\documentclass[aps,twocolumn]{revtex4}
\usepackage{graphicx}    \usepackage{rotating}    \usepackage{dcolumn}
\usepackage{epsfig} 

\begin{document}

\title{Transmission of Information in Active Networks} 

\author{M.  S.  Baptista} \affiliation{Max-Planck-Institut f\"ur Physik
Komplexer Systeme, N\"othnitzerstr. 38, D-01187 Dresden, Deutschland}
\author{J. Kurths} \affiliation{Universit{\"a}t Potsdam, Institut
  f{\"u}r Physik Am Neuen Palais 10, D-14469 Potsdam, Deutschland}
\date{\today}

\begin{abstract}
Shannon's Capacity Theorem is the main concept behind the Theory of
Communication. It says that if the amount of information contained in a signal
is smaller than the channel capacity of a physical media of communication, it
can be transmitted with arbitrarily small probability of error. This theorem
is usually applicable to ideal channels of communication in which the
information to be transmitted does not alter the passive characteristics of
the channel that basically tries to reproduce the source of information. For
an {\it active channel}, a network formed by elements that are dynamical
systems (such as neurons, chaotic or periodic oscillators), it is unclear if
such theorem is applicable, once an active channel can adapt to the input of a
signal, altering its capacity. To shed light into this matter, we show, among
other results, how to calculate the information capacity of an active channel
of communication. Then, we show that the {\it channel capacity} depends on
whether the active channel is self-excitable or not and that, contrary
to a current belief, desynchronization can provide an environment in which
large amounts of information can be transmitted in a channel that is
self-excitable. An interesting case of a self-excitable active channel
is a network of electrically connected Hindmarsh-Rose chaotic neurons.
\end{abstract}

\maketitle 


\section{Introduction}

Given an arbitrary time dependent stimulus that externally excites an active
channel, a network formed by systems that have some intrinsic dynamics
(e.g. oscillators and neurons), how much information from such stimulus can be
realized by measuring the time evolution of one of the elements of the channel
? The work in Ref.
\cite{roland0} shows that 50$\%$ of the information about light
displacements might be lost after being processed by the H1 neuron, sensitive
to image motion around a vertical axis, a neuron localized in a small neural
network of the Chrysomya magacephala fly , the lobula plate. Does that mean
that the H1 neuron has an information capacity lower than the information
contained in the light stimulus ?  Or does that mean that information is lost
due to the presence of internal noise ?

In order to be able to shed light into these questions, we need to know how to
calculate the information capacity of an active channel, and, for practical
purposes, to understand how an active channel (an active network composed by
dynamical systems) behaves if it is set up to operate with its information
capacity. As we shall see, there is a parameter route through which the
information capacity is reached, and this route can be established in terms of
either the coupling strengths or the level of synchronism (behavior) among the
elements forming the active channel. While information might not always be
easy to be measured or quantified in experiments, one might be able to state
about how good is an active channel to transmit information by measuring
synchronization, a phenomenom which is often not only possible to observe but
also relatively easy to characterize.

Synchronization is vital for modern methods of digital communication that rely
on the synchronous operation of many subsystems
\cite{kihara}. Similarly, transport networks depend crucially on 
the synchronous operation of each subnetwork. If one subnetwork gets out of
synchrony, the whole network might failure to function properly.  So, it would
be intuitive to say that complex systems should have subsystems that operate
in synchrony for a proper functioning. In fact, synchronization between
neurons in the brain is believed to provide a good environment for information
transmission.  This comes from a fundamental hypothesis of neurobiology
\cite{malsburg1,freeman,singer,gelade} that 
synchronization \cite{book_synchro,book_synchro1} functionally binds neural
networks coding the same feature or objects.  This hypothesis raised one of
the most important contemporary debates in neurobiology, but is still
controversial
\cite{pareti,shadlen} because desynchronization seems to play an
important role in the perception of objects as well. 

The analyses are carried out using among others two quantities suitable for
the treatment of information transmission in active channels, the {\it channel
capacity} and the {\it system capacity}. In short, the channel capacity
measures the maximum rate with which information is exchanged between two
elements of the active channel, a path along which information can flow
in the active channel. On the other hand, the system capacity is the
maximum of the Kolmogorov-Sinai entropy, the total amount of independent
information that can be simultaneously transmitted between all the pairs of
elements of the active channel.  

Among our main results, we show that the channel capacity of an 
active channel depends on whether the active channel is {\it self-excitable}
or not (see definition in Sec. \ref{capacities}).  Active channels composed of
non self-excitable systems (such as R\"ossler-type oscillators) achieve its
maximal channel capacity to transmit information whenever its elements
are in complete synchrony. On the other hand, active channels composed of
self-excitable systems (such as neurons), achieve its maximal channel
capacity when there is still at least one degree of freedom or characteristic
oscillation (time-scale) that is out of synchrony.  In the case of active
channels formed by spiking-bursting neurons (Hindmarsh-Rose electrically
coupled), the maximal channel capacity to transmit information is achieved
when the neurons phase synchronize in the slow time-scale (bursting) and
desynchronize in the fast time-scale (spiking). Thus, synchronization in
neural networks might be two folded.  Depending on the type of measurement
made, one can agree or disagree with the binding hypothesis.

Therefore, in this work, we build a bridge between Shannon's Theory of
Communication
\cite{shannon} and the Theory of Information in dynamical systems
\cite{ruelle} contributing to the development of a nonlinear Theory of
Communication applied to dynamical systems, shading some light in these
paradigms of neurobiology.  These new ideas, concepts, and theoretical
approaches unravel the relation between stimuli \cite{comment_stimuli},
information capacity, and synchronization, in a nonlinear media of
communication, the {\it active channel}, a network formed by elements that are
dynamical systems.

\section{MUTUAL INFORMATION RATE}

A good communication system as visualized by Shannon comprises a {\it
transmitter} who transforms a message in a signal suitable to be transmitted
through the channel and a {\it receiver} who recovers the message from the
signal.  In analogy to this definition and in order to properly deal with
information transmission in dynamical systems, we define three subspaces in
the active channel. The subspace $\alpha$ that generates a suitable message to
be transmitted, regarded as the transmitter (an element of the network), the
subspace $\beta$ where the message can be recovered, regarded as the receiver
(another element of the network), and finally the composed subspace
($\alpha,\beta$), considered to form one {\it communication channel}, an union
of the subspaces $\alpha$ and $\beta$, also a subspace of the active channel.
So, both the transmitter and the receiver belong to the communication
channel. Herein, the more is the information exchanged between the receiver
and the transmitter, the more information about the transmitter's trajectory
can be realized in the receiver. The trajectory of the transmitter (receiver)
represents the evolution in time of the transmitter's (receiver's) position in
the subspaces $\alpha$ ($\beta$).

According to Shannon, the amount of information that can be measured in the
receiver, $x^{(\beta)}$, about the transmitter of information, $x^{(\alpha)}$, is
given by
\begin{equation}
I(\alpha,\beta)=H^{(\alpha)}+H^{(\beta)}-H^{(\alpha,\beta)}, 
\label{I_S}
\end{equation}
\noindent
also known as mutual information between the transmitter and the
receiver. $H^{(\alpha)}$ is the information produced by the transmitter,
$H^{(\beta)}$, the one produced (or measured in) by the receiver, and
$H^{(\alpha,\beta)}$ the one produced in the composed subspace, also known as the
joint entropy between the receiver and the transmitter.  To calculate the
terms in Eq. (\ref{I_S}) for systems where events in the future are connected
to events in the past (systems with correlation), one usually needs to coarse
grain the domain of the subspaces $\alpha$ and $\beta$ into $n$ equal
size-$\epsilon$ \cite{comment_IS} intervals $\alpha_{i}$ and $\beta_{j}$, with
$i,j=1,\ldots,n$ ($n=1/\epsilon$), being that $P^{(\alpha)}_m$
($P^{(\beta)}_m$) represents the probability of an event, e.g. a trajectory
point in the subspace $\alpha$ ($\beta$) visiting a sequence of $L$
intervals. The trajectory remains a time $\tau$ in an interval $\epsilon$. If
one is working with maps, $\tau$=1. The term $P^{(\alpha,\beta)}_{m}$
represents the probability of a composed event, e.g. a trajectory point
visiting an itinerary following a sequence of $L$ areas, each area delimited
by the intervals $\alpha_i$ and $\beta_j$, as represented in Fig.
\ref{ac_fig01}. Then, in Eq.  (\ref{I_S}), $H^{(\alpha)}$=$-\sum_{m}
P^{(\alpha)}_m \ln P^{(\alpha)}_m$, $H^{(\beta)}$=$ -\sum_{m} P^{(\beta)}_m
\ln P^{(\beta)}_m$, and $H^{(\alpha,\beta)}$=-$\sum_{m} P^{(\alpha,\beta)}_{m}
\ln P^{(\alpha , \beta)}_{m}$, where we have taken the
limit of $(\epsilon, \tau)
\rightarrow 0, L \rightarrow
\infty$.

Notice that all the terms in Eq.  (\ref{I_S}) tend to infinity as $(\epsilon,
\tau) \rightarrow 0, L \rightarrow \infty$. So, it is advantageous to work with
terms $\sigma=H/(\tau L)$ that measure the amount of information per time
unit, which are finite quantities in the active channel. So, we rewrite Eq.
(\ref{I_S}) as
\begin{equation}
  I(\alpha,\beta) = \lim_{L \rightarrow \infty, (\tau, \epsilon)
    \rightarrow 0} \tau L (\sigma^{(\alpha)} + \sigma^{(\beta)} - \sigma^{(\alpha,\beta)})
  \label{I_S_finite} 
\end{equation}
\noindent
$I_C(\alpha,\beta) = I(\alpha,\beta)/(\tau L)$ is the {\it \bf mutual
information rate} (MIR) between the transmitter ($\alpha$) and the receiver
($\beta$). Based on the results of \cite{arnold}, the term
$\sigma^{(\alpha,\beta)}$ is the Kolmogorov-Sinai (KS) entropy \cite{pesin} of
the trajectory in the subspace $(\alpha,\beta)$ \cite{comment_IS}, regarded as
$H_{KS}^{(\alpha,\beta)}$.  Imagine that the receiver has a finite physical
coupling with the transmitter. From Takens theorem \cite{takens}, the entropy
of a trajectory calculated in a subspace, e.g. $(\alpha)$, should provide the
entropy of the trajectory in the whole space $(\alpha,\beta)$, which leads to
that $\sigma^{(\alpha)}$=$\sigma^{(\beta)}$=$\sigma^{(\alpha,\beta)}$, and
therefore, $I_C(\alpha,\beta)$=$H_{KS}^{(\alpha,\beta)}$. Independent on the
coupling strength and on the synchronization level between the receiver and
the transmitter, one arrives that the MIR is constant and given by
$H_{KS}^{(\alpha,\beta)}$. Naturally, in order for one to gain such an amount
of information rate, one might have to realize an infinite number of
observations in the receiver's trajectory and one has to have access to a good
trajectory projection (subspace). But, in communication, it is desirable that
information about the transmitter can be "instantaneously" realized in the
receiver. In addition, measurements performed in the subspace $\beta$ of
the receiver do not necessarely contain all the information content produced
by either the whole active channel or a communication channel. For these
reasons, we introduce a consistent definition for the MIR between two
subspaces (elements) in a network
\begin{equation}
I_C(\alpha,\beta) = D_1^{(\alpha)}|\lambda^{(\alpha)}| +
D_1^{(\beta)}|\lambda^{(\beta)}| - H_{KS}^{(\alpha,\beta)}
\label{I_C_ab}
\end{equation}
\noindent
where $\lambda^{(\alpha)}$ and $\lambda^{(\beta)}$ are the Lyapunov exponents
of the trajectories in the subspaces $\alpha$ and $\beta$, respectively, which
measures how nearby trajectories exponentially diverge in these subspaces,
$D_1^{(\alpha)}$ and $D_1^{(\beta)}$ are the information dimension of the
trajectory in these subspaces [see Appendix \ref{apendiceA}], and $|.|$ is the
modulus operation
\cite{comment_I_C_ab}. By doing that, we assume that the first two terms
($D_1^{(\alpha)}|\lambda^{(\alpha)}| + D_1^{(\beta)}|\lambda^{(\beta)}|$) in
the right side of Eq. (\ref{I_C_ab}) measure the information produced by both
the receiver and the transmitter as if they were uncoupled, i.e., uncorrelated
"random" variables (no phase-space reconstruction \cite{takens} is employed in
the measurable data from the subspaces $\alpha$ and $\beta$), and the last
term ($H_{KS}^{(\alpha,\beta)}$) provides the dependence between them. These
are actually the basic assumptions behind the definition of mutual information
provided by Shannon \cite{shannon} to random variables being transmitted
through a noisy channel. For more details about how to analytically calculate
the terms $\lambda^{(.)}$, see Appendixes \ref{apendiceA} and \ref{apendiceB}.

\section{Information capacity, excitability, and susceptibility}\label{capacities}

In order to study the way information is transmitted in active networks, we
introduce quantities and terminologies that assist us to better present our
ideas and approaches.

An {\bf active channel} is an active network constructed using $Q$ elements
that have some intrinsic dynamics and can be described by classical dynamical
systems, such as chaotic oscillators, neurons, phase oscillators, and so on.
Every pair of elements forms a {\bf communication channel} and the rate with
which information is exchanged between these elements, a transmitter and a
receiver, is given by the mutual information rate (MIR) between them.

The {\bf channel capacity}, $\mathcal{C}_C$, of a communication channel is
defined as the maximum of the MIR for this communication channel formed by a
pair of elements, the receiver and the transmitter, with respect to many
possible coupling strengths among the elements, for a given network
topology. Thus, the channel capacity is the maximal possible amount of
information that two elements within the network with a given topology can
exchange, a local measure that quantifies the point-to-point rate with which
information is being transmitted. Notice that a communication channel is a
subset of an active network.

The {\bf system capacity}, $\mathcal{C}_S$, of an active network composed by
$Q$ elements is defined as the maximum of the Kolmogorov-Sinai (KS) entropy,
$H_{KS}$, of the whole active network ($H_{KS}
\geq H^{(\alpha,\beta)}_{KS}$), with respect to many
possible coupling strengths among the elements, for a given network
topology. The Kolmogorov-Sinai entropy offers an appropriate way of obtaining
the entropy production of a dynamical system. Here, it provides a global
measure of how much information can be simultaneously transmitted among all
communication channels. Therefore, $\mathcal{C}_S$ bounds
$\mathcal{C}_C(\alpha,\beta)$ as well as the KS-entropy, $H_{KS}$, of an
active network, calculated for a given coupling strength, bounds the MIR
between two elements, $I_C(\alpha,\beta)$, calculated for the same coupling
strength. Thus,
\begin{eqnarray}
\mathcal{C}_C(\alpha,\beta) & \leq & \mathcal{C}_S \nonumber \\
I_C(\alpha,\beta) & \leq & H_{KS}
\label{limite} 
\end{eqnarray}

An active channel is said to be {\bf self-excitable} ({\bf not
self-excitable}) when $\mathcal{C}_C>H_{KS}^{(0)}$ (when $\mathcal{C}_C \leq
H_{KS}^{(0)}$), with $H_{KS}^{(0)}$ representing the KS entropy of one of the
$Q$ elements forming the active channel, before they are coupled. Analogously,
we can also define self-excitability in terms of the channel capacity. For a
self-excitable channel, it is true that $\mathcal{C}_S/Q>H_{KS}^{(0)}$. Thus,
in a self-excitable active channel $H_{KS}$ increases as the coupling strength
among the elements increases.

An active channel is said to be {\bf susceptible} if
$\mathcal{C}_C>H_{KS}^{(r)}$ and {\bf not susceptible} if $\mathcal{C}_C
\leq H_{KS}^{(r)}$, where $H_{KS}^{(r)}$ represents the KS entropy of the
uncoupled receiver. So, a susceptible channel does not resist the action of
the stimulus provided by the transmitter or the dynamical alteration caused by
the coupling configuration in the active channel. These alterations might
produce also a self-excitable channel. It is to be expected that a self-excitable
channel is also a susceptible one.

\section{THE CHAOTIC CHANNEL}

Here, we analyze how a source of information can be transmitted through a
channel that stretches the amplitude of the information signal
\cite{celso,murilo}.  The Lyapunov exponent of the receiver,
$\lambda^{\beta}$, is always positive even if there is no coupling between the
transmitter and the receiver. Part of the information transmitted might be
lost due to the presence of chaos in the channel.  We assume that a general
source of information can be modeled by a chaotic system.

\begin{figure}[!h]
  \centerline{\hbox{\psfig{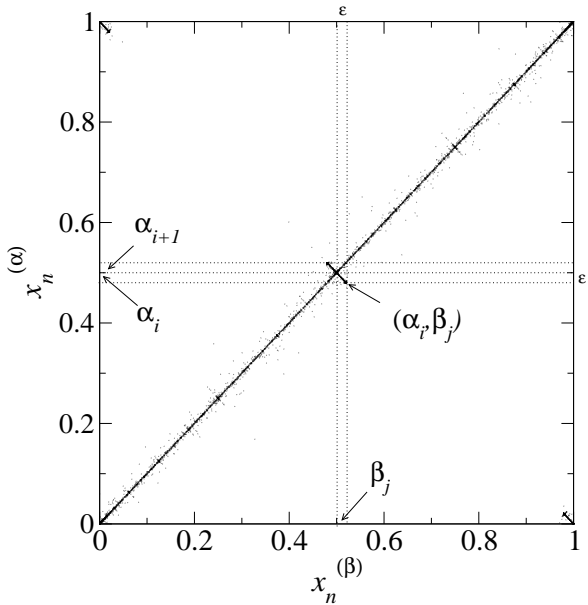}}}
  \caption{The subspace $\alpha$ of the transmitter who generates the information, the
  subspace $\beta$, where information about the transmitter can be realized,
and the composed subspace
  ($\alpha,\beta$) that represents a two-dimensional active channel of communication.}
  \label{ac_fig01}
\end{figure}

A model of the chaotic channel is given by two one-dimensional chaotic maps
bidirectionally coupled \cite{comment1}, 
\begin{eqnarray}
 x_{n+1}^{(\beta)} = 2x_n^{(\beta)} +
2c(x_n^{(\alpha)} - x_n^{(\beta)}), mod (1) \label{eq1} \\
x_{n+1}^{(\alpha)} = 2x_n^{(\alpha)} +
2c(x_n^{(\beta)}-x_n^{(\alpha)}), mod (1) \label{eq2}
\end{eqnarray}
\noindent
where the subspace of Eq. (\ref{eq1}) is regarded as the receiver and
Eq. (\ref{eq2}), the transmitter. In Fig. \ref{ac_fig01}, we represent
the map trajectory for $c=0.24$, the coupling strength. This map has
two Lyapunov exponents: $\lambda_1 = \ln{(2)}$ and $\lambda_2 =
\ln{(2-4c)}$, $\lambda_1$ measures the exponential divergence of
nearby trajectories in the direction of the synchronization manifold
defined as $x^{(\alpha)} - x^{(\beta)}$=0, and $\lambda_2$ the
exponential divergence of nearby trajectories in the direction
transversal to the synchronization manifold.  The exponents
$\lambda^{(\alpha)}$ and $\lambda^{(\beta)}$ that measure the
exponential divergence of trajectories along the subspaces $\alpha$
and $\beta$ are equal to $\max{(\lambda_1,\lambda_2)}$=$\lambda_1$,
and therefore, $\lambda^{(\alpha)}=\lambda^{(\beta)}$=$\lambda_1$,
since the maps have equal parameters [see Sec. \ref{secaoVI}]. One can
also arrive to this result by noting that the Lyapunov exponent of a
typical 1D projection of a 2D chaotic set (with two positive Lyapunov
exponents) is given by the Largest exponent. Since the trajectories in
the subspaces $\alpha$ and $\beta$ have uniform probability
distribution and the information dimension of the trajectory in the
composed subspace is $D_1$=2, a one dimensional projection of it
should provide $D_1^{(\alpha)}$=$D_1^{(\beta)}$=1 (see Appendixes
\ref{apendiceA} and \ref{apendiceB}). Using a result by Pesin
\cite{pesin}, the Kolmogorov-Sinai entropy of a chaotic system is the
summation of all the positive Lyapunov exponents, $H_{KS} = \sum^+
\lambda$.  For a two-dimensional channel, $H_{KS}^{(\alpha,\beta)}$
equals the entropy of the whole channel, $H_{KS}$. So,
$H_{KS}=\lambda_1+\lambda_2$, if $\lambda_2 \geq 0$, or
$H_{KS}=\lambda_1$, otherwise. Therefore, we arrive that the rate with
which information can be retrieved in the receiver about the stimulus
generated in the transmitter is given by $I_C = \lambda_1 -
\lambda_2$, if $\lambda_2 \geq 0$, or $I_C=\lambda_1$, otherwise. So,
$\mathcal{C}_C=\ln{(2)}$ and $\mathcal{C}_S=2\ln{(2)}$. An increase in
the coupling leads to an increase in the MIR and a decrease in
$H_{KS}$.

To relate the MIR with the synchronization level of this chaotic channel, we
make a convenient coordinate transformation into new variables $x^{\parallel}$
and $x^{\perp}$ (see Appendix \ref{apendiceA}) such that the exponential divergence on
$x^{\perp}$ is minimal ($x^{\perp}$ is oriented along the contracting
direction) and $x^{\parallel}$ is orthogonal to $x^{\perp}$ ($x^{\parallel}$
is oriented along the expanding direction).  Such a transformation, for the
systems to be treated here, is $x^{(\alpha
\beta)\parallel}_n$=$x_n^{(\alpha)}+x_n^{(\beta)}$ and $x^{(\alpha
\beta)\perp}_n$=$x_n^{(\alpha)}-x_n^{(\beta)}$, with the synchronization manifold
given by $x^{\perp}_n$=0. The mapping in this new coordinate generates the
conditional Lyapunov exponents $\lambda^{\parallel}=\lambda_1$ and
$\lambda^{\perp}=\lambda_2$. In contrast to the conditional exponents
defined in Ref. \cite{pecora}, the obtained
exponents produce physically consistent quantities (ergodic invariants)
\cite{mendes} even for situations when complete synchronization is absent. The
transformed equations of motion in these new variables provide not only the
same Lyapunov exponents but also, for this example, the same KS entropy ($H_{KS}
= \lambda^{\parallel} +
\lambda^{\perp}$, if $\lambda^{\perp} \geq 0$, or  $H_{KS}
=\lambda^{\parallel}$, otherwise) of the original equations, and
advantageously supply us with a way to understand the synchronization level
between the two subsystems. $\lambda^{\parallel} >
\lambda^{\perp}$, and we recover the conjecture of Ref.  \cite{murilo}, $I_C$ =
$\lambda^{\parallel} -
\lambda^{\perp}$, if $\lambda^{\perp} \geq 0$ or $I_C=\lambda^{\parallel}$,
otherwise. This conjecture provides an easy way of solving Eq.  (\ref{I_C_ab})
for an active channel linking information to synchronization. The more
synchronization, the smaller is $x^{\perp}$, and therefore, nearby initial
conditions in this variable will diverge exponentially in a smaller rate,
i.e., the conditional exponent associated with this variable, $\lambda^{\perp}$,
is smaller. If $\lambda^{\perp}$ is very small, $\lambda^{\parallel}$ can be
associated with the amount of information production of the synchronous
trajectories (between the transmitter and receiver), otherwise, it is
associated with the excitation of the channel. The more excitation in the
channel the larger is $\lambda^{\parallel}$. So, to achieve larger amounts of
information transmission it is required that either the excitation or
the synchronization level are large, or both.  In this channel, as we increase
the coupling, $H_{KS}$ decreases due to an increase in the synchronization
level, which leads to an increase in the MIR. $\mathcal{C}_C=\min{(H_{KS})}
\neq
\mathcal{C}_S$ is achieved for a configuration when the synchronization is
maximal. {\it Therefore, the larger the coupling is, the less information the
whole active channel produces (KS-entropy, $H_{KS}$), but the larger the MIR
between a receiver and a transmiter is, which means that the more information
about the transmitter can be measured in the receiver. More synchronization
implies more information transmission. This channel is not self-excitable and
since $\mathcal{C}_C=H_{KS}^{(r)}$, it is not susceptible, because its
capacity is limited by the capacity of the receiver to generate information}

In order to calculate the MIR of a communication channel in a large chaotic
active channel, we need to use the coordinate transformation $x^{\parallel}$
and $x^{\perp}$.  This transformation enables one to calculate the MIR between
two subsystems as if they were detached from the active channel.  Imagine an
active channel formed by $Q$ fully bidirectionally coupled chaotic systems:
\begin{equation}
x^{(j)}_{n+1} = 2x^{(j)}_{n} + \sum_{i=1}^Q 2c(x^{(i)}_n-x^{(j)}_n)
mod (1)
\end{equation}
\noindent
with $j=[1,\ldots,Q]$.  Now, we can define $[Q \times (Q-1)]/2$ pairs
of subspaces.  For instance, the pair of subspaces formed by the subsystem
$x^{(1)}$ and the subsystem $x^{(2)}$, with
$x^{(12)\parallel}_n$=$x_n^{(1)}+x_n^{(2)}$ and
$x^{(12)\perp}_n$=$x_n^{(1)}-x_n^{(2)}$.  Any pair of subspaces produces the
same conditional exponents $\lambda^{\parallel} =
\ln{(2)}$ and $\lambda^{\perp} = \ln{|2(1-Qc)|}$ 
[see Appendix \ref{apendiceA}].  In fact, this system produces one Lyapunov
exponent $\lambda =
\lambda^{\parallel}$ and $(Q-1)$ equal others $\lambda =
\lambda^{\perp}$, and so, our defined conditional exponents can be
related to the Lyapunov exponents even in higher-dimensional systems.  So, the
MIR [see Eq.  (\ref{I_C})] between any two subsystems $x^{(k)}$ and $x^{(l)}$
is $I_C(x^{(k)},x^{(l)}) = -\ln{(1-Qc)}$ bits per iteration of the mapping,
for $c \leq 1/(2Q)$ ($\lambda^{\perp} \geq 0$). If there is no coupling
($c$=0), then $I_C(x^{(k)},x^{(l)}) = 0$ and no information is exchanged
between both subspaces. For $c \leq 1/(2Q)$, the larger is $c$ the more
synchronous a transmitter, say $x^{(k)}$, is with a receiver, say $x^{(l)}$,
and the more information is exchanged. The channel capacity for all
communication channels is achieved for $c
\geq 1/(2Q)$, when $I_C(x^{(k)},x^{(l)})=\ln{(2)}=\min{(H_{KS})}$, and the
network completely synchronizes ($\lambda^{\perp} < 0$). {\it This type of
active channel is not self-excitable. Notice that the introduction of one more
element into this channel \cite{comment_stimuli} does not alter
$\mathcal{C}_C$. It is also not susceptible.}

\section{THE PERIODIC CHANNEL}

The purpose of the present work is to describe how information is transmitted
via an active media, a network formed by dynamical systems. There are three
possible asymptotic stable behaviors for an autonomous dynamical system:
chaotic, periodic, or quasi-periodic. A quasi-periodic behavior can be usually
replaced by either a chaotic or a periodic one, by an arbitrary
perturbation. For that reason, we neglect such a state and focus the attention
on active channels that are either chaotic or periodic. 

The purpose of the present section is dedicated to analyze how a source of
information can be transmitted through active channels that are periodic,
channels that squeeze the amplitude of the information signal.  More
specifically, channels whose receiver behaves in a periodic fashion (its
Lyapunov exponent, $\lambda^{\beta}$, is negative).

\begin{figure}[]
\centerline{\hbox{\psfig{file=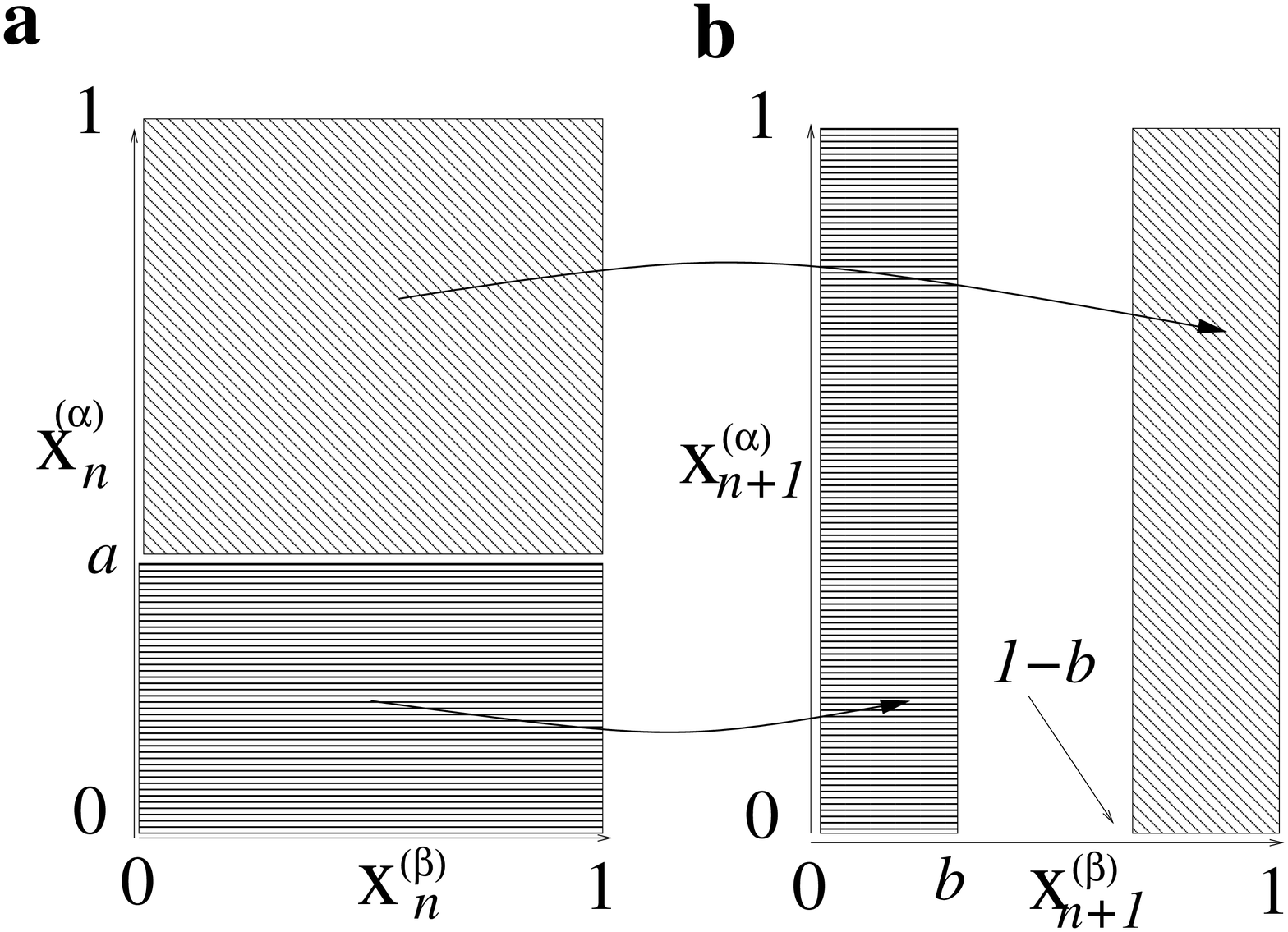,height=8cm,width=8cm}}}
\caption{{\bf a}, The baker's map squeezes the rectangle above $a$ in {\bf a}
    into the right rectangle in {\bf b} and it squeezes the rectangle bellow
    $a$ in {\bf a} into the left rectangle in {\bf b}. Both rectangles in {\bf
    a} are stretched in the horizontal direction. From {\bf a} to {\bf b}, we
    represent one application of the generalized baker's map.}
\label{ac_fig2}
\end{figure}

\begin{figure}[]
\centerline{\hbox{\psfig{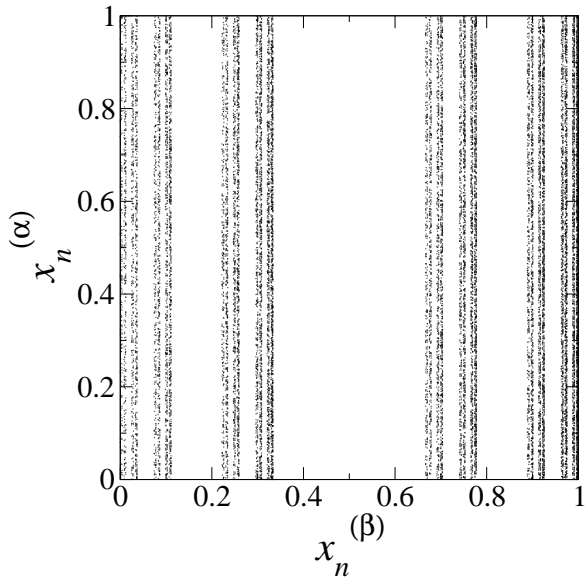}}}
\caption{The asymptotic solution of the map in Fig. \ref{ac_fig2}, formed by a series of vertical
    strips forming a fractal set in the horizontal direction.}
\label{ac_fig2_1}
\end{figure}

It is to be expected that in the periodic channel a fractal set appears, when
$\lambda_1 \leq |\lambda_2|$ (assuming a bidimensional channel). This clearly
imposes severe limits for the recovery of information in the receiver.  The
periodic channel can be imagined as a filter. As shown in
\cite{badii}, chaotic signals being transmitted through filters might produce
an output with higher dimension due to the appearance of a fractal set.  To
see that we study the generalized baker's map \cite{ott}, shown in Figs.
\ref{ac_fig2}({\bf a}-{\bf b}) and in Fig. \ref{ac_fig2_1}. All the information produced in the transmitter
$x^{(\alpha)}$ is transfered to the receiver $x^{(\beta)}$, but with a time
delay. To make it more clear, note that in Fig. \ref{ac_fig2}{\bf b}, by
recognizing if the received signal is either smaller or larger than $b$ at the
iteration time $n+1$, one is able to know if the position of the transmitter
was lower or higher than $a$, at the iteration time $n$. By looking the
received signal at higher resolution one is able to predict with higher
resolution the position of the transmitter farer away in the past.

In order to calculate the MIR between the transmitter and the receiver using
Eq. (\ref{I_C_ab}) note that $\lambda^{(\alpha)}$=$a\ln{[1/a]} +
(1-a)\ln{[1/(1-a)]}$, $D_1^{(\alpha)}$=1, $\lambda^{(\beta)}$=$a\ln{(b)} +
(1-a)\ln{(b)}$, and $D_1^{(\beta)}$=$\lambda^{(\alpha)}/|\lambda^{(\beta)}|$
(with $D_1^{(\beta)}<1$) and
$\lambda^{(\alpha,\beta)}=H_{KS}=\lambda^{(\alpha)}+\lambda^{(\beta)}$. If
$D_1^{(\beta)}<1$, $\lambda^{(\beta)} < 0$, and $\lambda^{(\alpha)} \leq
|\lambda^{(\beta)}|$, and a fractal set takes place.  As demonstrated in
Ref. \cite{ledrappier}, the information content of this fractal set is
$D_1^{(\beta)} |\lambda^{(\beta)}|$
\cite{comment3}. Thus, $I_C$=$\lambda^{(\alpha)} + D_1^{(\beta)}|\lambda^{(\beta)}| -
H_{KS}$, and we arrive at $I_C$=$\lambda^{(\alpha)}$ per iteration, since
$H_{KS}=\lambda^{(\alpha)}$ and
$|\lambda^{(\beta)}|D_1^{(\beta)}=\lambda^{(\alpha)}$. $H_{KS}=\lambda^{(\alpha)}$
because the fractal set does not contribute to the KS entropy of the full
chaotic map.  {\it So, $\mathcal{C}_C$ depends on the amount of information
produced in the transmitter, a typical characteristic of a susceptible
channel. Unlike the chaotic channel that is robust to small noise intensities
\cite{mu21}, in the periodic channel $\mathcal{C}_C$ might be extreme
sensitive to noise of even arbitrary amplitudes.}


In active channels, the receiver might influence the transmitter behavior. We
can imagine a bidirectional coupling scheme for which a periodic uncoupled
receiver might make an uncoupled chaotic transmitter to behave also
periodically, after the coupling is switched on. This type of periodic channel
is thus not susceptible to adapt to stimuli, and $\mathcal{C}_C$=0.

The periodic channel might be relevant to understand the role of subthreshold
oscillations in the processing of information. These oscillations are observed
in the brain, in particular to regions associated with motor reaction and
learning like the Inferior Olive \cite{olive}.  They appear because groups of
neurons interact in such a way that the potential in the neurons membrane is
not sufficient to induce a spike.  Usually, the oscillations are reduced to a
limit cycle, a periodic behavior.

\section{THE NEURON CHANNEL}

We illustrate our ideas in a relevant type of excitable active
channel, the chaotic neural channel \cite{abarbanel,tito,roland}, formed by a
network of electrically connected Hindmarsh-Rose (HR) neuron models
\cite{HR}. This network possesses both characteristics of the periodic and
chaotic channels, since it has both positive and negative Lyapunov exponents. A
fractal set that occupies a small portion of the phase space coexists
with a chaotic set that occupies most of the phase space. Due to the negative
exponents, the dynamics is strongly compressed along the stable directions,
the stable manifolds. The result is that the observable dynamics of the
neurons lies along the unstable manifolds, and thus, the negative exponents do
not contribute to the recovered information. This is a consequence of the
Sinai-Ruelle-Bowen (SRB) assumption.

Each pair of neurons can be treated as an active channel of communication, one
neuron $\alpha_k$ performing the transmitter task and the other $\beta_l$ the
receiver. In Ref. \cite{murilo}, we have stated that the amount of information
in one chaotic channel should be always smaller than the information produced
by the network, $H_{KS}$. Thus, $I_C(\alpha_k,\beta_l) \leq H_{KS}$ [see
Eq. (\ref{limite})].  By working with large networks, composed of many
elements, we should expect that the same information travels simultaneously
along many different channels.  This property, often desired in networks,
makes it a reliable medium for information transmission because it introduces
in the network a large amount of redundancy, which results into a sum of all
the MIR in the communication channels larger than $H_{KS}$.  Even if one or
many channels are blocked, the information still finds its destination.  So,
to treat networks composed by $Q$ chaotic systems, we expect that {\small
\begin{equation}
\langle I_C \rangle \leq H_{KS}. 
\label{I_C_H_KS}
\end{equation}}
where $\langle I_C \rangle=\sum_{k,l} I_{C}(\alpha_k,\beta_l)/P$ represents
the average amount of MIR of the whole network, $P$ is the number of
communication channels given by $P=[Q(Q-1)]/2$, and $Q$ is the number of
neurons. For the neuron channel, we consider that $\mathcal{C}_C$ represents
the maximal of $\langle I_C \rangle$ for many coupling configurations. The
average amount of redundancy in the network is defined to be $\langle R
\rangle =
\frac{\langle I_C \rangle}{H_{KS}}$, and thus, if the network is completely
synchronized $\langle R \rangle =$1, and if the network is completely
uncoupled $\langle R \rangle$=0.

We consider a network composed of $Q$=4 bidirectionally fully coupled neurons:
\begin{eqnarray}
\dot{x}_i &=& y_i + 3x_i^2 - x_i^3-z_i + I_i +
\sum_j A_{ji}(x_j-x_i) \nonumber \\
\dot{y}_i &=& 1-5x_i^2-y_i \label{HR} \\
\dot{z}_i  &=& -rz_i + 4r(x_i+1.6) \nonumber
\end{eqnarray}
\noindent
The parameter that modulates the slow dynamics is set to $r$=0.005, such that
each neuron is chaotic.  $i$ and $j$, with $j\neq i$ assume
values within the set $[1,\ldots,Q]$.  $\alpha_k$ represents the subsystem of
the variables $(x_k,y_k,z_k)$ and $\beta_l$ represents the subsystem of the
variables $(x_l,y_l,z_l)$, where $k$=$[1,\ldots,Q-1]$ and
$l$=$[k+1,\ldots,Q]$.  $A_{ji}=A_{ij}$
\cite{comment1} is the strength of the electrical coupling between
the neurons represented by $\alpha_j$ and $\alpha_i$.  The external stimulus
$I_1$, in $\alpha_1$, is set to be equal to $I_1$ = 3.25-$\delta I$, and then,
$I_2$=$I_1$+$\delta I$, $I_3$=$I_1$-$\delta I$, $I_4$=$I_1$+$\delta I$, with
$\delta I$ = 0.00001.  Initial conditions are $x$=-1.3078+$\eta$,
$y$=-7.3218+$\eta$, and $z$=3.3530+$\eta$, where $\eta$ is an uniform random
number within [0,0.001].

Four synchronization phenomena are relevant to be considered [see Appendixes
\ref{apendiceA} and \ref{apendiceC}]. Bursting phase synchronization (BPS),
when at least one pair of neurons is phase synchronous in the bursts, partial
phase synchronization (PPS) when at least one pair of neurons is phase
synchronous in the bursts and in the spikes, phase synchronization (PS), when
all the pairs of neurons are phase synchronous in the bursts and in the
spikes, and complete synchronization (CS).  An evidence for the presence of
bursting or spiking phase synchronization is found if the condition
$\max{(\Delta N^n)}/P \leq 1$ is satisfied, where $\Delta N^n = \sum_{k,l}
|N_k^n - N_l^n|$ and $N_k^n$ represents the number of spikes/bursts in
$\alpha_k$, at the time the neuron $\alpha_1$ suffered its $n$-th
spike/burst. This condition is threshold dependent, but it will be employed
here for the purposes of illustration.

\begin{widetext}

\begin{figure}[!h]
\centerline{\hbox{\psfig{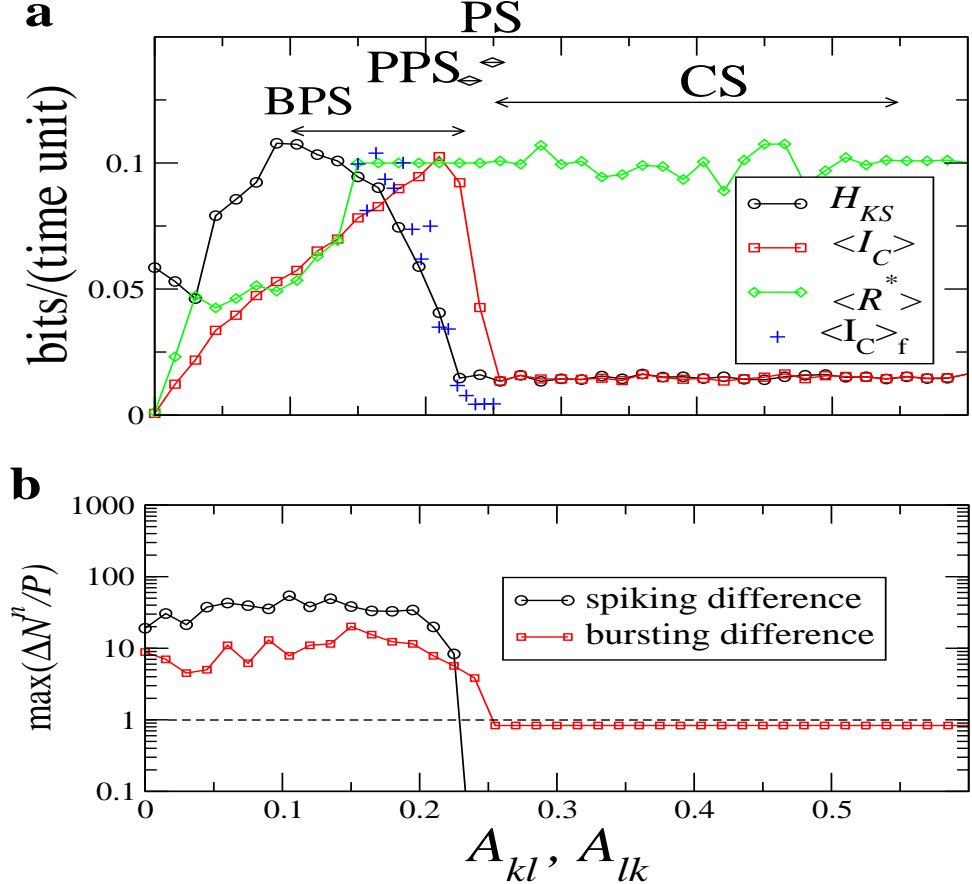}}} 
\caption{[Color online] {\bf a}, Circles, squares, and diamonds show $H_{KS}$, $\langle
  I_C \rangle$, and $\langle R^* \rangle$=$\langle R \rangle/10$ for a network
  of 4 fully connected neurons. The necessary conditions for the proper
  calculation of $\langle I_C \rangle$ are not satisfied for coupling
  strengths that produce stronger types of PS ($A_{kl}=[0.15,0.25]$), see Appendix
  \ref{apendiceA}. There, $\langle I_C \rangle$ should be estimated by a
  finite time MIR, indicated by $\langle I_C \rangle_f$. For this coupling
  strength interval, we set $\langle R^* \rangle$=1. {\bf b}, Circles and
  squares represent the maximal average spiking and bursting difference,
  $\max{(\Delta N^n/P)}$, after $n$=200 bursts, in a log vertical axis.
  Values of $\max{(\Delta N^n/P)}$ smaller than the dashed line are an
  evidence that there is PS. The upper arrows in {\bf a} indicate the coupling
  strength intervals for which we find BPS ($A_{kl} \cong [0.1,0.23[$),
  PPS ($A_{kl}$=$[0.23,0.245[$), PS ($A_{kl}$=$[0.245,0.25[$), and CS ($A_{kl}
  \geq 0.25$). To obtain the information in units of bits we divide the
  related equations by $\ln{(2)}$. $I_C$ is calculated using Eq.  (\ref{I_C}),
  which produces similar values to the ones obtained by calculating the MIR
  from the entropy and joint entropy of symbolic sequences generated from the
  trajectory of pair of neurons.}
\label{paper_HR_fig0}
\end{figure}

\end{widetext}

This example, illustrated by Fig. \ref{paper_HR_fig0}, shows three fundamental
characteristics of an active channel: 

{\it (i) Excitation enhances $H_{KS}$ and the MIR of the communication
channels.}  With no coupling, the rate of information production in each
neuron is approximately $H_{KS}^{(0)} \cong 0.014$ and $\langle I_C \rangle$
is null. For a coupling strength of $A_{kl,lk}>0.01$, each pair of neuron
exchanges (in average) information with each other in a rate larger than the
individual rate with no coupling. So, an increase in the coupling strength is
simultaneously followed by an increase in both $H_{KS}$ and the rate of
information production of each individual neuron, resulting in an increase of
$\langle I_C \rangle$, meaning also an increase in the MIR of the
communication channels, a typical characteristic of a self-excitable channel.

{\it (ii) Synchronization does not necessarily mean high levels of information
transmission. } When the network reaches its system capacity ($A_{kl}
\approx$0.08), the spikes and the bursts are highly desynchronous [Fig.
\ref{paper_HR_fig0}{\bf b}] by usual definitions of phase synchronization
[see Appendix \ref{apendiceC}], but both $\langle I_C
\rangle$ and
the redundancy $\langle R \rangle$ are high.  At this point, we have to
remember that MIR means the amount of excitation minus the amount of
desynchronization. The amount of excitation is of the order of the
maximal Lyapunov exponent of the network, which is large, since the network is
excited, much larger than the amount of desynchronization. On the other hand,
for $A_{kl} \geq 0.23$, when the neurons phase (PPS or PS) or completely
synchronize ($A_{kl} \geq 0.23$), $\langle I_C
\rangle$ abruptly drops approaching the low value of $H_{KS}$, much lower than
$\mathcal{C}_S$, as if the whole network were formed by one single neuron.
The redundancy is high but few information can be transmitted in the network.

{\it (iii) BPS provides an ideal environment for information
transmission.} When BPS is present, $\langle I_C \rangle$ and the
redundancy are high.  $H_{KS} \cong
\langle I_C \rangle$.  That suggests that BPS plays
an important role in the reliable exchange of information that demands rapid
responses and a large amount of information transmission. Each neuron
maintains a high level of independent activity (given by the desynchronous
spikes) and simultaneously a moderate level of synchrony (synchronism in the
bursts) that allows a neuron to "talk" to another. These characteristics
are usually desirable in sensory neurons \cite{neural_code} and in the ones
responsible for motor reaction processes.

\section{ACTIVE CHANNELS FORMED BY NON-EQUAL ELEMENTS}\label{secaoVI}

Here, we briefly describe the dependence of the MIR on the parameter
mismatches between the elements forming an active channel.  For such a case,
$\lambda^{(\alpha)}$ and $\lambda^{(\beta)}$ typically differ, in
Eq. (\ref{I_C_ab}). As a consequence, the channel capacity is lower than when
the parameters do not mismatch and Eq. (\ref{I_C}) should be considered as an
upper bound (it might overestimate the real value) for the MIR between the
subspaces $\alpha_k$ and $\alpha_l$
\cite{explica_cc}. Also, a parameter mismatch might enhance the value of the
MIR if the coupling is kept constant while the parameters are changed. For
simplicity, let us represent an active channel formed by two coupled
unidimensional maps by ($\vec{x}^{(\alpha)}_{n+1},
\vec{x}^{(\beta)}_{n+1}$). Thus, $\lambda^{(\alpha)}$, which  gives how nearby
trajectories exponentially diverge along the subspace $\alpha$, can be
calculated from $\frac{\partial{\vec{x}^{(\alpha)}_{n+1}}}
{\partial{\vec{x}^{(\alpha)}_{n}}} +
\frac{\partial{\vec{x}^{(\alpha)}_{n+1}}} {\partial{\vec{x}^{(\beta)}_{n}}}$, the
sum of the terms in the row of the Jacobian with respect to subsystem $\alpha$,
and $\lambda^{(\beta)}$ from $\frac{\partial{\vec{x}^{(\beta)}_{n+1}}}
{\partial{\vec{x}^{(\alpha)}_{n}}} +
\frac{\partial{\vec{x}^{(\beta)}_{n+1}}} {\partial{\vec{x}^{(\beta)}_{n}}}$, the
sum of the terms in the row of the Jacobian with respect to subsystem $\beta$.

As an illustration, imagine the following chaotic channel: 
\begin{eqnarray}
x_{n+1}^{(\beta)}
= (a-\epsilon)x_n^{(\beta)} + 2c(x_n^{(\alpha)} - x_n^{(\beta)}),  mod (1)
\nonumber \\
x_{n+1}^{(\alpha)} = ax_n^{(\alpha)} + 2c(x_n^{(\beta)}-x_n^{(\alpha)}), 
mod (1) 
\end{eqnarray}
\noindent
with $a>1$ and $\epsilon$ is the parameter mismatch. Then, we arrive
at $\lambda^{(\alpha)}$=$\ln{(a)}$, $\lambda^{(\beta)}$=$\ln{(|a-\epsilon|)}$,
$\lambda_1$=$\lambda^{\parallel}$= $\ln{(|\mu_1+\mu_2|)}$, and
$\lambda_2$=$\lambda^{\perp}$= $\ln{(|\mu_1-\mu_2|)}$, with
$\mu_1=a-2c-\epsilon/2$ and $\mu_2=2\sqrt{(c^2+\epsilon^2/16)}$.  If there is
no coupling ($c$=0), and even if $\epsilon \neq 0$, $I_C(\alpha,\beta)=0$. To
see that, we assume $\epsilon\rightarrow$0, which lead us to
$\lambda_1=\ln{(|a-\epsilon/2|)}$ and $\lambda_2=\ln{(|a-4c-\epsilon/2|)}$,
and then we use the expansion $\ln{(a-\epsilon)}$=$\ln{(a)}-\epsilon/2$, in
the exponents. The larger is $\epsilon$, the smaller is the channel capacity,
$\mathcal{C}_C$, which is reached for parameters that lead to $\lambda_2=0$
($a-4c-\epsilon/2=1$), when the maps, although not completely synchronous,
have close trajectories. If $\lambda_2>0$ ($c\leq -1/4-\epsilon/8+a/4$), then
$I_C(c,\epsilon>0) \geq I_C(c,\epsilon=0)$, which means that a parameter
mismatch can enhance the MIR of the channel, more precisely by a maximal
amount of $\epsilon/4$. That means that parameter mismatches for sufficiently
small coupling strengths enhance the synchronization level in an active
channel.  Notice that Eq. (\ref{I_C}) is indeed an upper bound for the MIR,
being that the difference between $I_C$ obtained from Eq. (\ref{I_C}) and the
one from Eq. (\ref{I_C_ab}) is of the order of $\epsilon^2$, for small
$\epsilon$ and $c$, and for larger $c$ it is smaller than $\epsilon/4$.  If
$\epsilon$=0, then $\lambda^{(\alpha)}$=$\lambda^{(\beta)}$=$\ln{(a)}$, which
results in $\mathcal{C}_C$=$\ln{(a)}$, the maximal possible value for the MIR.
Finally, the system capacity, $\mathcal{C}_S=2\ln{(a)}$, is reached when $c$=0
and $\epsilon=0$ \cite{comment_fractal}. {\it This channel is not susceptible
for $\epsilon=0$. If $\epsilon \neq 0$, it becomes susceptible, since
$\mathcal{C}_C>H_{KS}^{(r)}=\ln{(|a-\epsilon|)}$. However, independently on
$\epsilon$, this channel is not self-excitable.}

For larger HR neural networks (up to 50 neurons) described by Eq. (\ref{HR}), the
system capacity is reached for a network of equal neurons, but with a non-zero
coupling strength.  We find that $\mathcal{C}_S$ is reached for a Small World
network geometry
\cite{strogatz}, being that $\mathcal{C}_S$ increases
linearly with the number of neurons, $Q$, by $\mathcal{C}_S$ =13.75$Q$
bits/burst. The larger $Q$ is, the smaller the coupling strengths, which
are considered to be equal. Also, the average number of connections, $\gamma$,
that each neuron receives scales linearly with the number of neurons as
$\gamma \propto 0.5 Q$. 

Networks formed by HR neurons connected to their nearest neighbors, forming a
ring, can be regarded as models for the small networks of electrically
connected neurons found in the Inferior Olive \cite{japao} that regulates the
transmission of information between the cerebellum and the cortex, and is
responsible for motor control and learning. In this type of network, we find
that the system capacity increases linearly with the number of neurons by
$\mathcal{C}_S$ =9.5$N$ bits/burst, being achieved always for the same small
coupling strength (~0.08).  Therefore, the capacity does not depend on 
the coupling strength. Network configurations for which the system capacity is
reached operate also with a large MIR in each communication channel. This is
an optimal configuration for networks found in the Inferior Olive that demand
large amounts of information transmission for an efficient cerebellar
learning. Naturally, we do not expect that the neural networks found in the
Inferior Olive are formed by equal neurons. So, the calculated channel and
system capacities should be interpreted as an upper bound for these quantities
in realistic models of the networks found in there.

\section{THE MUTUAL INFORMATION RATE BETWEEN OTHER SUBSPACES}

In this work, we are mainly interested in calculating the MIR between the
subspaces $\alpha$ and $\beta$ [Eqs. (\ref{I_C_ab}) and (\ref{I_C})]. That
means that we are mainly interested in knowing the rate of information that
can be realized from a transmitter (one element of the active channel) by
measuring the signal of a receiver (another element of the active channel).

However, it is of general interest to learn how to calculate the MIR between
groups of elements or between different subspaces of the active channel. For
example, in many experimental situations, one cannot obtain the signal of an
isolated element but rather an average field, or average quantity, such as the
quantity $\vec{x}^{\parallel}$, which can be imagined as an average field
($\vec{x}^{\parallel}_{kl} = \vec{x}_{k} + \vec{x}_{l}$) between two elements.

As briefly described here, in fact, we can also calculate the MIR between
other subspaces. As an illustration, we consider the calculation of the MIR
between the subspaces defined by the coordinate transformations
$\vec{x}^{\parallel}$ and $\vec{x}^{\perp}$, regarded as
$I_C(\vec{x}^{\parallel},\vec{x}^{\perp})$.

Typically, we should expect that $I_C(\vec{x}^{\parallel},\vec{x}^{\perp})
\neq I_C(\alpha,\beta)$, and therefore, the MIR is coordinate dependent. While
$I_C(\alpha,\beta)$ measures the rate with which information about the
transmitter $\alpha$ can be realized by observing the receiver $\beta$,
$I_C(\vec{x}^{\parallel},\vec{x}^{\perp})$ measures the rate with which
information about $\vec{x}^{\parallel}$ can be realized by observing
$\vec{x}^{\perp}$. Naturally, at a situation when CS takes place, nothing
about $\vec{x}^{\parallel}$ can be said by observing $\vec{x}^{\perp}$, since
$\vec{x}^{\perp}$=0, and therefore,
$I_C(\vec{x}^{\parallel},\vec{x}^{\perp})$=0. So, to know from
$I_C(\vec{x}^{\parallel},\vec{x}^{\perp})$ if one element exchanges high
amounts of information with another element, not only
$I_C(\vec{x}^{\parallel},\vec{x}^{\perp})$ should be low, but also
$\lambda^{\perp}$.

Notice that if the network is composed by elements with a linear dynamics (or
piecewise linear for continuous dynamics), we can always find a linear
transformation from which $I_C(\vec{x}^{\parallel},\vec{x}^{\perp})$ can be
calculated from the state variables $\vec{x}$, or $I_C(\alpha,\beta)$ from the
transformed variables $\vec{x}^{\parallel}$ and $\vec{x}^{\perp}$. Usually,
one would calculate $I_C(\vec{x}^{\parallel},\vec{x}^{\perp})$ using the
variables $\vec{x}^{\parallel}$ and $\vec{x}^{\perp}$ and $I_C(\alpha,\beta)$
using the variables $\vec{x}^{\alpha}$ and $\vec{x}^{\beta}$.

Therefore, even if a measurement cannot provide the signal of isolated
elements in an active network, there might be situations in which we can still
calculate the mutual information rate exchanged between two isolate elements
from the measurements of averages.

\section{THE TIME DEPENDENT (NOISY) ACTIVE CHANNEL}

If all the elements forming the active channel suffer the influence of
the same time dependent stimulus (or noise) and both the stimulus is
completely uncorrelated with respect to the variables
$\vec{x}^{\perp}$ and $\vec{x}^{\parallel}$ and also $\vec{x}^{\perp}
\rightarrow 0$ [such that the Jacobian in Eq. (\ref{jacobian}) is
approximately block diagonal], then both exponents $\lambda^{\perp}$ and
$\lambda^{\parallel}$, calculated for the autonomous channel, are not modified
by the introduction of the stimulus. Also, if the largest Lyapunov exponent of
the channel is not affected by the introduction of the stimulus,
$\lambda^{\parallel}$ is also not modified. Thus, if the previous conditions
are satisfied, Eq.  (\ref{I_C}) calculated for the autonomous channel gives
the upper bound for the mutual information of the non-autonomous channel.

The periodic channel fully preserves the transmitted information,
however it might transform it in a fractal set which is vulnerable to
noise. So, the information rate recovered in the receiver can be
sensitively dependent on the noise level in the channel. Chaotic
channels tend to destroy part of the information transmitted, even
without the presence of noise ($\sum \lambda^{\perp}>0$).  However,
they might offer a nice way to deal with additive noise. As shown in
Ref.  \cite{mu21}, Gaussian noise with small variance added to a
chaotic trajectory can be completely filtered out.

The action of more general types of time dependent stimulus that alters the
dynamics of the active channel still needs better clarification.


\section{CONCLUSION}

An {\it active channel} is an active network composed by dynamical
systems. Every pair of elements forms a {\it communication channel} and the
rate with which information is exchanged between two elements, a transmitter
and a receiver, is given by the mutual information rate (MIR) between
them. The maximum rate of information that can be transmitted in a
communication channel of an active channel is regarded as the {\it channel
capacity}, $\mathcal{C}_C$, and the maximum rate of information produced in
the whole active channel is regarded as {\it system capacity}, the maximum of
the Kolmogorov-Sinai entropy, $H_{KS}$, of the active channel. All these
maximums are calculated with respect to many possible coupling strengths among
the elements, for a given network topology.

We can organize the active channels in two main classes. Periodic or
chaotic. A chaotic (periodic) channel is composed by a receiver that behaves
in a chaotic (periodic) fashion, for long time intervals. Chaotic channels
formed by HR neuron networks are self-excitable, which means that the channel
capacity, $\mathcal{C}_C$, is larger than $H_{KS}^0$, the Kolmogorov-Sinai
entropy of all the elements forming the active channel when they are
uncoupled. In a self-excitable channel, a transmitter and a receiver (together
with all the other elements forming the channel) mutually increase their
capacity for information production, leading to an increase in their channel
capacity.  So, the introduction of stimuli in a self-excitable active channel
might increase its channel capacity for information transmission. Not all
chaotic channels present self-excitability. For example, we have not verified
such a property in active channels formed by linearly coupled R\"ossler
oscillators or by linearly coupled Chua's circuit (while in the Double Scroll
attractor regime). It is to be expected that a periodic channel is non
self-excitable.

More synchronization results in an increase of the MIR between two elements in
an active channel, regarded as transmitter and receiver, if as the transmitter
becomes more synchronous with the receiver, simultaneously also the
Kolmogorov-Sinai entropy decreases, meaning that synchronization is
accompanied by a reduction of the excitation in the channel. This situation is
to be expected in non self-excitable channels. In self-excitable channels, a
large amount of information transmission can be obtained when the bursts are
phase synchronous while the spikes are highly desynchronous.

Periodic channels might allow the complete transmission of the information
signal provided by the transmitter. On the other hand, the transmitted
information in a chaotic channel might be lost due to the presence of
non-synchronous chaotic trajectories, if the transmitter is weakly coupled to
the receiver. However, while a periodic channel might be very sensible to the
presence of even arbitrarily small noise intensities, chaotic channels might
be robust.


\begin{appendix} 

\section{Conditional exponents and complete synchronization (CS)}\label{apendiceA}

Assume $\vec{x}_k$ to describe the state variables of subsystem $k$.  The
parallel and perpendicular subspaces are defined to be a transformation in the
variables of the subsystems that maximize the calculated mutual
information. For the cases here studied, the parallel subspace between
$\alpha_k$ and $\alpha_l$ is defined as $\vec{x}^{\parallel}_{kl}$ =
$\vec{x}_k+\vec{x}_l$, and the transversal subspace is defined as
$\vec{x}^{\perp}_{kl}$ = $\vec{x}_k-\vec{x}_l$. For a network of $Q$ elements,
$k=[1,Q-1]$ and $l=[k+1,Q]$. Writing the equations of motion in these new
variables, one can separate the transformed equations into subsystems that
contain only terms of that subsystem. So, a network of $Q$ elements formed by
systems of dimension $m$ ($\mathcal{R}^{m}$), can be broken down in $Q(Q-1)/2$
subsystems of dimensionality $2m$ ($\mathcal{R}^{2m}$). Then, the conditional
exponents of the neural network between two subspaces measure the exponential
divergence of nearby trajectories of the transformed equations for these two
subspaces, which in practice is calculated using the following Jacobian in the
method of Ref.\cite{ruelle} 
\begin{widetext}
{\LARGE
\begin{equation}
\begin{array}{cc}


  \frac{\partial \dot{\vec{x}}_{kl}^{\perp}}{\partial \vec{x}_{kl}^{\perp}} & 
  \frac{\partial \dot{\vec{x}}_{kl}^{\perp}}{\partial \vec{x}_{kl}^{\parallel}} \\
  \frac{\partial \dot{\vec{x}}_{kl}^{\parallel}}{\partial \vec{x}_{kl}^{\perp}} & 
  \frac{\partial \dot{\vec{x}}_{kl}^{\parallel}}{\partial \vec{x}_{kl}^{\parallel}}
\end{array}
\label{jacobian}
\end{equation}
}
\noindent
which is 
{\large
\[
\begin{array}{cccccc}
  3x^{\parallel}-\frac{3(x^{\perp2}+x^{\parallel2})}{4}+F &
  1 &
  -1   &
  3x^{\perp} - \frac{3x^{\parallel}x^{\perp}}{2}  &
  0   &
  0 \\

  -5x^{\parallel} &
  -1 &
  0 &
  -5x^{\perp} &
  0 &
  0 \\

4r &
 0 &
 -r &
 0 &
0 &
 0 \\

  {3x^{\perp}} - \frac{3x^{\perp}x^{\parallel}}{2} + G&
 0 &
 0 &
   {3x^{\parallel}} - 
\frac{3(x^{\parallel2} + x^{\perp2})}{4}& 
 1 &
-1 \\

 -{5x^{\perp}} &
 0 &
 0 &
  -{5x^{\parallel}}   &
 -1 &
 0 \\

 0 &
 0 &
 0 &
 4r &
 0 &
 -r
\end{array}
\]
}
\end{widetext}
\noindent
where $x$ stands for $x_{kl}$. For a network of $Q$ fully connected neurons
with equal coupling strengths, $F=-QA_{kl}$. When
$\xi=[(Q-2)/2]\vec{x}_{ij}^{\parallel} - \sum_{k=1}^Q \vec{x}_{k}$ (with $k \neq
i,j$) is either orders of magnitude smaller than the quadratic terms (no
synchronization) or $\xi$=0 (CS), then $G\cong 0$. 
For parameter values close to or when PPS or PS is present, the quadratic terms are
also small and $G$ cannot be neglected. In Fig. \ref{paper_HR_fig0}, this
happens for the parameter region $A_{kl},A_{lk} \cong [0.15,0.25]$. There,
Eq. (\ref{I_C_H_KS}) seems to be violated. To resolve that, we force $\xi
\cong 0$, which leads to $G \cong 0$, condition for which the Jacobian in
Eq. (\ref{jacobian}) can be used. We set the initial conditions all equal, and
integrate the system for a small time interval (5 bursts which is equivalent
to about 50 spikes) to estimate a finite time averaged MIR,
indicated in this figure by $\langle I_C
\rangle_f$.  Within the time scale for
which phenomena happen in real biological neural networks, periodic state as
well as chaotic state in the asymptotic sense might never be observed, but
rather a transient state whose subspaces $\vec{x}^{\parallel}$ and
$\vec{x}^{\perp}$ possess finite time conditional exponents.  Finite time
quantities are well defined in dynamical systems.

For $Q$=2, $F=-2A_{kl}$ and $G=0$. When CS takes place in a network formed by
$Q$ neurons, the only term of the Jacobian that changes is $F$. For this
Jacobian, we can calculate that CS appears if $F(Q)\leq F(Q=2)$, where
$F(Q=2)=-2A_{kl}(Q=2)$, with $A_{kl}(Q=2)$=0.5 being the coupling for which
complete synchronization appears for a configuration of two coupled neurons
[see also \cite{hasler}].  So, the coupling to reach CS is $A_{kl}
\geq 1/Q$, when the second largest Lyapunov as well as all transversal
conditional exponents are negative. At this point, the trajectory distance
between any pair of neurons tends to zero.

For the active channels composed by coupled one-dimensional maps
$$
x^{(j)}_{n+1} = 2x^{(j)}_{n} + \sum_{i=1}^Q 2c(x_n^{(i)}-x_n^{(j)}), mod
(1), 
$$
\noindent
we can calculate the mutual information in each communication channel exactly,
with no need of any special conditions.  The network equations can be broken
down in subspaces that depend only on the parallel or transversal variables of
that subspace.  So, $x^{(kl)\perp}_{n+1}$=$Fx_n^{(kl)\perp}$, and
$x^{(kl)\parallel}_{n+1}$=$Gx_n^{(kl)\parallel} +H(x_n^{(kl)\perp})$. $H$ does
not participate in the calculation of the conditional exponents and can be
ignored. For a fully connected network formed by $Q$ maps with equal coupling
strengths $c$, $F$=$2(1-Qc)$ and $G=2$.  The conditional exponents are
$\lambda^{\perp}$=$\ln{(|F|)}$ and $\lambda^{\parallel}=\ln{(|G|)}$.  For a
network of $Q$=4 maps bidirectionally connected to their nearest neighbors
forming a closed ring, i.e.  $x^{(i)}$ is connected to $x^{(i+1)}$ and to
$x^{(i-1)}$, and $x^{(Q)}$ is connected to $x^{(Q-1)}$ and to $x^{(1)}$, then
$F$=$2[1-2c]$ ($F$=$2[1-c]$) and $G$=2 for subspaces whose pair of systems
have a direct connection (no connection).  This network completely
synchronizes when all the $\lambda^{\perp}<$0, and thus, when $c>1/2$.

\section{Information dimension, Lyapunov exponents and MIR}\label{apendiceB}

We consider an active channel formed by only one communication channel,
composed by two coupled unidimensional systems which produce an attractor
$\Gamma$ with at most two positive Lyapunov exponents. $\Gamma$ is
corse-grained with volumes of size $\epsilon$
\cite{comment_IS}, and for $\epsilon \rightarrow 0$, we have that
$\sum_{i} P_i^{\prime} \ln{P_i^{\prime}}/\ln{(\epsilon)} = D_1$, with $D_1$
being the information dimension of $\Gamma$, a quantity that measures the
information content of $\Gamma$ and $P_i^{\prime}$ is the probability of
finding a trajectory point in one of the $i$ volumes of size $\epsilon$.  The
average probability $\langle P \rangle$ of finding a trajectory following an
itinerary visiting one of the possible $m$ combinations of sequences of $L$
volumes of size $\epsilon$ for a time interval $L\tau$ is $\langle P \rangle
\propto \exp^{- \tau L\sum_j^+ D_1^{(j)} \bullet
\lambda_j}$, and 
$D_1^{(j)}$ ($\sum_j D_1^{(j)} = D_1$ and $D_1^{(j)} \in [0,1]$) are the
partial information dimensions
\cite{grassberger},  a quantity that measures
the information content of $\Gamma$ along the direction $j$, either parallel
or orthogonal to the trajectory, and $\lambda_j$ are the Lyapunov exponents in
the direction $j$. $\bullet$ is the inner product. Assuming that the
distribution of trajectory points is smooth along unstable directions
(associated with positive exponents) and the system possesses a
Sinai-Ruelle-Bowen (SRB) measure
\cite{ruelle}, $D_1^{(j)} \rightarrow$1 if $\lambda_j>0$. From
\cite{pesin,ledrappier}, $1/(\tau L)\sum_m P_i
\ln{P_i} = H_{KS}$ for SRB systems. To understand how that is derived, 
we assume uniformity in the probability distribution, $-\sum_m P_i
\ln{P_i} = -\ln{\langle P \rangle}$.  Then, the term $H_{KS}^{(\alpha,\beta)}$
[in Eq. (\ref{I_C_ab})] can be calculated by knowing that $-\ln{\langle P
\rangle/(\tau L)} \sim \sum_j^+ \lambda_j$.  
Now, we make the intuitive hypothesis that the terms $\sigma^{(\alpha)}$ and
$\sigma^{(\beta)}$, in Eq. (\ref{I_S_finite}), preserve the physical
quantities used to calculate $\sigma^{(\alpha,\beta)}$. So, if
$\sigma^{(\alpha,\beta)}$ is a function of the information dimension and the
Lyapunov exponents of the trajectory on the subspaces ($\alpha,\beta$),
similarly, $\sigma^{(\alpha)}$ and $\sigma^{(\beta)}$ should be a function of
these quantities. This hypothesis provides the terms $D_1^{(\alpha)}
|\lambda^{(\alpha)}| + D_1^{(\beta)} |\lambda^{(\beta)}|$ in
Eq. (\ref{I_C_ab}). If a projection of $\Gamma$ onto the lower-dimensional
subspaces $\alpha$ and $\beta$ produces a fractal set, one has to consider the
absolute value of the negative Lyapunov exponent \cite{ledrappier}. Otherwise,
$D_1^{(\alpha)}=D_1^{(\beta)}=\max{[D_1^{(j)}]}$, assuming that the subspaces
$\alpha$ and $\beta$ contain only the dynamics of the expanding directions,
i.e., either they are unidimensional or they can be reduced to a
unidimensional subspace by a mapping of the flow \cite{comment_I_C_ab}. For
most of the chaotic channels here studied, $\max{[D_1^{(j)}]}$=1, for typical
projections of $\Gamma$. Further, we calculate the Lyapunov exponents of the
transmitter and the receiver as if they were detached from each other. If the
elements that compose the channel have equal parameters,
$\lambda^{(\alpha)}=\lambda^{(\beta)}$=$\lambda_1$=$\lambda^{\parallel}$,
otherwise, $\lambda^{(\alpha)} \neq \lambda^{(\beta)}$.

For the considered networks, 
each pair of elements, in the transformed variables of the parallel and
transversal subspaces, forms a composed subspace $(\alpha_k,\beta_l)$ that
produces at most two positive conditional exponents. The previous analysis
applies for each communication channel of this network, since each pair of
neurons has a dynamics equivalent to a bidimensional discrete map with at
most two positive conditional Lyapunov exponents, a larger one
$\lambda^{\parallel}_{kl}$ and a smaller one $\lambda^{\perp}_{kl}$. Then,
\begin{equation}
  I_C(\alpha_k,\beta_l) = \lambda^{\parallel}_{kl} - \lambda^{\perp}_{kl},
\label{I_C}
\end{equation}
\noindent
if $\lambda_{kl}^{\perp}>0$, or
$I_C(\alpha_k,\beta_l)=\lambda^{\parallel}_{kl}$, otherwise.  To derive Eq.
(\ref{I_C}), we have used that $\lambda^{(\alpha_k)}$=$\lambda^{(\beta_l)}$=
$\lambda^{\parallel}_{kl}$, $D_1^{(.)}=\max{(D_1^{(j)})}$=1, and
$H_{KS}^{(\alpha_k,\beta_l)}$ is the sum of all positive conditional exponents
of the composed subspace $(\alpha_k,\beta_l)$. This can be done whenever the
subspace $(\alpha_k,\beta_l)$ is separable from the whole network. If the
elements forming the channel have different parameters and one still wants to
use Eq. (\ref{I_C}), be in mind that such an equation might provide an upper
bound for the MIR between these two subspaces.


For arbitrary networks, for simplicity let us imagine two coupled oscillators
$\vec{x}^{(\alpha)}$ and $\vec{x}^{(\beta)}$, coupled by a term $c$, the terms
$\lambda^{(\alpha)}$ and $\lambda^{(\beta)}$ from Eq. (\ref{I_C_ab}) can be
analytically or semi-analytically calculated if the coupling is either
sufficiently small (such that the elements forming the network are almost
decoupled) or sufficiently large (such that the whole network has a high level
of synchrony). At this situation, $\lambda^{(\alpha)}$ ($\lambda^{(\beta)}$)
is the Lyapunov exponent in the subspace of the oscillator
$\vec{x}^{(\alpha)}$ ($\vec{x}^{(\beta)}$), i.e., assume $\vec{x}^{(\beta)}=0$
($\vec{x}^{(\alpha)}=0$) and $c$=0, and then calculate the Lyapunov exponents
by the usual methods.  For such coupling strengths, this is an equivalent
approach to the one described in Sec. \ref{secaoVI}. For other coupling
strengths, we expect that $D_1^{(\alpha)} |\lambda^{(\alpha)}| + D_1^{(\beta)}
|\lambda^{(\beta)}| - H_{KS}^{(\alpha,\beta)} \leq
H_{KS}^{(\alpha,\beta)}$. If not, that points to the existence of trajectory
foldings in the lower-dimensional subspaces $\alpha$ and $\beta$ which results
in an overestimation of the Lyapunov exponents, $\lambda^{(\alpha)}$ and
$\lambda^{(\beta)}$. In such cases, $D_1^{(.)}$ should be underestimated
\cite{brian}, so balancing the action of the trajectory foldings. 

If the trajectory is very close to the synchronization manifold and so,
$\vec{x}_1 \cong \vec{x}_2 \cong
\ldots \cong
\vec{x}_Q$, the term 
$\partial \dot{\vec{x}}_{kl}^{\parallel}/\partial
\vec{x}_{kl}^{\parallel}$ in Eq. (\ref{jacobian}) gives the largest Lyapunov
exponent $\lambda$ of the network which equals the largest exponent of one
neuron, and thus, Eq. (\ref{I_C}) can be estimated by $I_C(\alpha_k,\alpha_l)
= \lambda - \lambda^{\perp}_{kl}$. This equation agrees with the intuitive
idea that the amount of information exchanged between two systems within a
large network is given by the amount of information production of one system
($\lambda$) minus the error in the transmission between both systems
($\lambda^{\perp}_{kl}$).


\section{Phase and Phase synchronization}\label{apendiceC}
Phase synchronization \cite{book_synchro} is a phenomenon defined by
\begin{equation}
|\phi_k - m \phi_l| \leq r, 
\label{phase_synchronization}
\end{equation}
\noindent
where $\phi_k$ and $\phi_l$ are the phases of two neurons $\alpha_k$ and
$\alpha_l$, $m=\omega_l/\omega_k$ is a real number \cite{murilo_irrational},
and $\omega_k$ and $\omega_l$ are the average frequencies of
oscillation of the neurons $\alpha_k$ and $\alpha_l$, and $r$ is a finite
number
\cite{baptista:2006}. In this work, we have used in
Eq. (\ref{phase_synchronization}) $m=1$, which means that we search for
$\omega_k:\omega_l$=1:1 (rational) phase synchronization
\cite{book_synchro}. If another type of $\omega_k:\omega_l$-PS is present, the
methods in Refs. \cite{baptista:2005,tiago:2007,baptista:2006} can detect.

The phase $\phi$ is a function constructed on a 2D subspace, whose trajectory
projection has proper rotation, i.e, it rotates around a well defined center
of rotation. So, the phase is a function of a subspace. Usually, a good 2D
subspace of the HR neurons is formed by the variables $x$
and $y$, and whenever there is proper rotation in this subspace the phase can
be calculated by \cite{tiago_PLA2007}
\begin{equation}
\phi(t)=\int_0^t \frac{\dot{y}x-\dot{x}y}{{(x^2+y^2)}}dt.
\label{phase_xy}
\end{equation} 
If there is no proper rotation in the subspace $(x,y)$ one can still find proper
rotation in the velocity subspace $(\dot{x},\dot{y})$ and a phase can be defined by
\cite{tiago_PLA2007}
\begin{equation}
\phi(t)=\int_0^t \frac{\ddot{y}\dot{x}-\ddot{x}\dot{y}}{{(\dot{x}^2+\dot{y}^2)}}dt.
\label{phase_dxdy}
\end{equation} 
If a good 2D subspace can be found, one can also define a phase by
means of Hilbert transformation, which basically transforms an
oscillatory scalar signal into a two components signal \cite{gabor}.
In the active channel of Eq. (\ref{HR}), for the coupling strength
interval $A_{kl} \cong [0,0.05]$, the subspace $(x,y)$ has proper
rotation, and therefore, phase is well defined and can be calculated
by Eq.  (\ref{phase_xy}). However, for this coupling interval, Eq.
(\ref{phase_synchronization}) is not satisfied, and therefore, there
is no PS between any pair of neurons in the subspace $(x,y)$.

For the coupling strength interval $A_{kl} \cong [0.05,0.23]$, the
neurons trajectories lose proper rotation both in the subspaces
$(x,y)$ and $(\dot{x},\dot{y})$.  The phase cannot be calculated
by Eq. (\ref{phase_xy}) or by Eq. (\ref{phase_dxdy}).  That
is due to the fact that the chaotic trajectory gets arbitrarily close
to the neighborhood of the equilibrium point $(x,y)$=$(0,0)$, a
manifestation that a homoclinic orbit to this point exists.  

In fact, the Hilbert transformation also fails to provide the phase from
either scalar signals $x$ or $y$, since these signals do not present any
longer an oscillatory behavior close to the equilibrium point. In such cases,
even the traditional technique to detect PS by defining the phase as a
function that grows 2$\pi$, whenever a trajectory component crosses a
threshold cannot be used. Since the trajectory comes arbitrarily close to the
equilibrium point, no threshold can be defined such that the phase difference
between pairs of neurons is bounded. Notice that by this definition the phase
difference equals $2\pi \Delta N$. For that reason, Fig.
\ref{paper_HR_fig0}{\bf b} would remain roughly as it is even if the
thresholds that define a spike and a burst are modified or even if
another variable (either $y$ or $z$) is used.  In this figure, a
burst (spike) in a neuron $\alpha_k$ is considered to start/end if
$x_k$ crosses the threshold defined by $x_k$=-1.0 ($x_k$=0.0).

In order to check if PS indeed exists in at least one subspace,
alternative methods of detection can be employed as proposed in Refs.
\cite{baptista:2005,tiago:2007}.  In short, if PS exists in a 
subspace then by observing one neuron trajectory at the time the other bursts
or spikes (or any typical event), there exists at least one special curve,
$\Gamma$, in this subspace, for which the points obtained from these
conditional observations do not visit the neighborhood of $\Gamma$. A curve
$\Gamma$ is defined in the following way. Given a point $x_0$ in the attractor
projected on the subspace of one neuron where the phase is defined, $\Gamma$
is the union of all points for which the phase, calculated from this initial
point $x_0$ reaches $n
\langle r
\rangle$, with
$n=1,2,3,\ldots,\infty$ and $\langle r \rangle$ a constant, usually
2$\pi$. So, note that an infinite number of curves $\Gamma$ can be
defined. For coupled systems with sufficiently close parameters that present
in some subspace proper rotation, if the points obtained from the
conditional observations do not visit the whole attractor projection on this
subspace, one can always find a curve $\Gamma$ that is far away from the
conditional observations. Therefore, for such cases, to state the existence of
PS one just has to check if the conditional observations are localized with
respect to the attractor projection on the subspace where the phase is
calculated.

\begin{figure}
\centerline{\hbox{\psfig{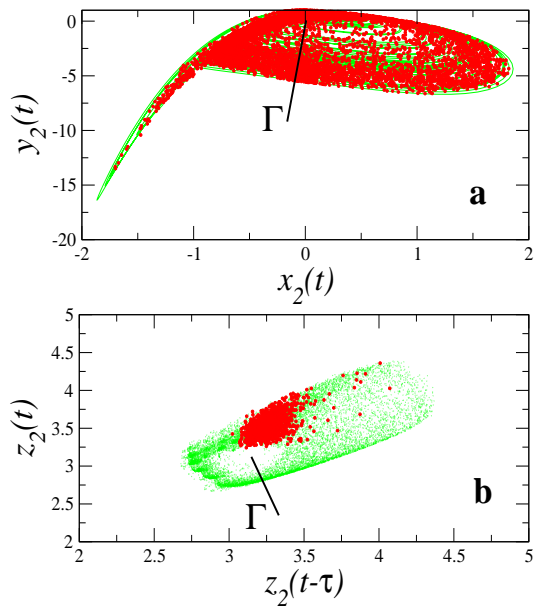}}} 
\caption{[Color online] The network of Eqs. (\ref{HR}) for 
$A_{kl}$=0.1. The curve $\Gamma$, a continuous curve transversal to the
trajectory, is pictorially represented by the straight line $\Gamma$.  {\bf
a}, the conditional observations are not localized and thus there is no PS in
this subspace. The light gray line (green online) represents the attractor
projection on the subspace $(x,y)$ of the neuron $\alpha_2$, and filled gray
circles (red online) represent the points obtained from the conditional
observations of the neuron $\alpha_2$ whenever the neuron $\alpha_4$
spikes. The point $(x,y)=(0.0)$ does not belong to $\Gamma$. {\bf b}, the
conditional observations are localized and thus there is PS in this
subspace. Light gray dots (green online) represent the reconstructed attractor
$z_2(t) \times z_2(t - \tau)$, for $\tau$=30, and filled circles (red online)
represent the points obtained from the conditional observation of neuron
$\alpha_2$, whenever the reconstructed trajectory of the neuron $\alpha_4$
crosses the threshold line $z_4(t-\tau)=3.25$ and $z_4(t)>3$.}
\label{fig_revised00}
\end{figure}

Conditional observations of the neuron trajectory $\alpha_k$ in the subspace
($x,y$), whenever another neuron $\alpha_l$ spikes, in the system modeled by
Eqs.  (\ref{HR}), are not localized with respect to a curve $\Gamma$, for the
coupling strength $0.05 < A_{kl} < 0.23$. An example can be seen in
Fig. \ref{fig_revised00}{\bf a}, for $A_{kl}=0.1$. The set of points produced
by the conditional observations are represented by dark gray circles (red
online), and the attractor by the light gray points (green online). Therefore,
there is no PS in the subspace $(x,y)$. However, the points obtained from the
conditional observations do not visit the whole attractor in the subspace
($x,y$). This is an evidence that there is PS in some other subspace.

In order to know on which subspace PS occurs, we proceed in the following way.
We reconstruct the neuron attractors by means of the time-delay technique,
using the variable $z$. This variable describes the slow time-scale,
responsible for the occurrence of bursts.  The reconstructed attractor $z(t)
\times z(t - \tau)$ has proper rotation (see Fig. \ref{fig_revised00}{\bf
b}) and the points obtained from the conditional observations do not visit the
neighborhood of a curve $\Gamma$, then, there is PS in this subspace.  Indeed,
we find localized sets with respect to a curve $\Gamma$ in the system of
Eqs. (\ref{HR}), in the reconstructed subspace ($z(t)
\times z(t - \tau)$),  for $A_{kl} \geq 0.1$.

So, for the parameter interval $A_{kl}=[0.1,23[$, there is no PS in the subspace
$(x,y)$ but there is PS in the subspace of the variable $z$.  In this type of
synchronous behavior, the bursts are phase synchronized while the spikes are
not.  This behavior is regarded as bursting phase synchronization (BPS).

For simplicity in the analysis, we say that BPS happens when at least one pair
of neurons is phase synchronous in the bursts.  Partial phase synchronization
(PPS) happens in the network when it is true that for at least one pair of
neurons Eq. (\ref{phase_synchronization}) is satisfied by the phases as
defined by either Eq. (\ref{phase_xy}) or Eq. (\ref{phase_dxdy}). In addition,
at the coupling strengths for which PPS appears, one positive Lyapunov and one
positive transversal conditional exponent become negative. At the coupling
strengths for which Eq. (\ref{phase_synchronization}) is satisfied for all
pair of neurons (there is PS), the second largest Lyapunov exponent and all
the transversal conditional exponents become non positive.

Notice that these phenomena happen in a hierarchical way organized by the
"intensity" of synchronization. The presence of a stronger type of
synchronization implies in the presence of other softer types of
synchronization in the following order: CS $\rightarrow$ PS $\rightarrow$ PPS
$\rightarrow$ BPS.

\section{RECOVERY OF INFORMATION}

Equations (\ref{I_S_finite}) and (\ref{I_C}) give the amount of information
that can be retrieved in the receiver per time unit. Imagine the neural
network. If, in average, a burst happens for a time interval $\Delta T$, one
can retrieve in the receiver an amount $I_C \Delta T$ of information about the
transmitter {\it per burst}. It is often desirable to known how much
information one single observation with precision $\epsilon$ can
provide. Assuming that observations are taken over in time intervals not
smaller than $\delta t = -\ln{(\epsilon)}/
\lambda^{\parallel}$, the maximal amount of information,
$I_m$, that can be retrieved in the receiver about the transmitter in each
observation is estimated by
$I_m=(\lambda^{\perp}/\lambda^{\parallel}-1)\ln{(\epsilon)}$.  We arrive at
this result by assuming $\delta t$ to be the memory time of the channel, the
time interval for which observations in the receiver trajectory with precision
$\epsilon$, at a time $t_0$, will provide no information about the transmitter
trajectory, at the time $t_0+\delta t$, and $I_m=I_C
\times \delta t$.

\section{TRANSIENT DYNAMICS}

If an active channel is being externally stimulated or if the initial
conditions are far away from the asymptotic (for large time intervals) stable
state (periodic or chaotic behavior), the channel will present a transient
dynamics. In such a case, Eqs. (\ref{I_C_ab}) and (\ref{I_C}) should remain
valid by the use of finite time conditional or Lyapunov exponents (assuming
$D_1^{j}$=1). As an illustrative example, an active channel that has an
asymptotic chaotic attractor might have to be treated as a periodic channel
(space contracting dynamics), if the initial conditions are far away from the
chaotic set and the dynamics is dominated by the stable directions.

\end{appendix}

\textbf{Acknowledgment} We thank R. Kliegl who has, by a series of
discussions, made us aware of the works in experimental psychology about the
capacity limit of short-term memory, D. Gauthier who has convinced us of the
importance to understanding how signals can be transmitted through channels
that are space contracting, H. Kantz for the conceptual discussions concerning
Eq. (\ref{I_C_ab}), K. Josi\'c, and T. Nishikawa for another important series of
discussions, and T. Pereira for a critical reading of Appendix C.

\end{document}